\documentclass[a4paper,11pt]{article}
\pdfoutput=1
\usepackage{jheppub}
\usepackage{textcase}
\usepackage{graphicx,epsfig,psfrag,amssymb,hyperref}
\usepackage{multirow}
\usepackage{color,graphicx,epsfig,psfrag,amsmath,empheq}
\usepackage[dvipsnames]{xcolor}
\usepackage{bm}
\usepackage{mathrsfs,amsfonts,soul,color}
\usepackage[caption=false]{subfig}
\usepackage{hepunits}
\usepackage{enumitem}
\usepackage{placeins}
\usepackage{textcomp}
\usepackage{tcolorbox}
\usepackage{float}
\usepackage[toc,page]{appendix}
\usepackage[ruled,vlined]{algorithm2e}
\usepackage{wrapfig}

\definecolor{darkblue}{rgb}{0.1,0.1,1.0}

\DeclareRobustCommand{\Eq}[1]{Eq.~(\ref{eq:#1})}
\DeclareRobustCommand{\Eqs}[2]{Eqs.~(\ref{eq:#1}) and (\ref{eq:#2})}

\DeclareRobustCommand{\Ref}[1]{Ref.~\cite{#1}}

\begin{document}


\title{Presenting Unbinned Differential Cross Section Results}

\author[a,b]{Miguel Arratia,}
\author[c]{Anja Butter,}
\author[d]{Mario Campanelli,}
\author[e]{Vincent Croft,}
\author[f]{Dag Gillberg,}
\author[g,h]{Aishik Ghosh,}
\author[i]{Kristin Lohwasser,}
\author[j]{Bogdan Malaescu,}
\author[k]{Vinicius Mikuni,}
\author[h,l]{Benjamin Nachman,}
\author[m,n]{Juan Rojo,}
\author[o,p]{Jesse Thaler,}
\author[q]{Ramon Winterhalder}

\affiliation[a]{Department of Physics and Astronomy, University of California, Riverside, CA 92521, USA}
\affiliation[b]{Thomas Jefferson National Accelerator Facility, Newport News, VA 23606, USA}
\affiliation[c]{Institut f\"ur Theoretische Physik, Universit\"at Heidelberg, Heidelberg, Germany}
\affiliation[d]{University College London, London WC1E 6BT, UK}
\affiliation[e]{Department of Physics and Astronomy, Tufts University, Boston, MA 02155, USA}
\affiliation[f]{Department of Physics, Carleton University, Ottawa ON K1S 5B6, Canada}
\affiliation[g]{Department of Physics and Astronomy, University of California, Irvine, CA 92697, USA}
\affiliation[h]{Physics Division, Lawrence Berkeley National Laboratory, Berkeley, CA 94720, USA}
\affiliation[i]{University of Sheffield, Sheffield, S10 2TN, UK}
\affiliation[j]{LPNHE, Sorbonne Universit\'e, Universit\'e de Paris, CNRS/IN2P3, Paris, France}
\affiliation[k]{National Energy Research Scientific Computing Center, Berkeley, CA 94720, USA}
\affiliation[l]{Berkeley Institute for Data Science, University of California, Berkeley, CA 94720, USA}
\affiliation[m]{Nikhef Theory Group, Science Park 105, 1098 XG Amsterdam, The Netherlands}
\affiliation[n]{Department of Physics and Astronomy, Vrije Universiteit Amsterdam,
NL-1081 HV Amsterdam, The Netherlands}
\affiliation[o]{Center for Theoretical Physics, Massachusetts Institute of Technology, Cambridge, MA 02139, USA}
\affiliation[p]{The NSF AI Institute for Artificial Intelligence and Fundamental Interactions}
\affiliation[q]{Centre for Cosmology, Particle Physics and Phenomenology (CP3), Universit\'e catholique de Louvain, 1348 Louvain-la-Neuve, Belgium}

\emailAdd{bpnachman@lbl.gov}

\abstract{
Machine learning tools have empowered a qualitatively new way to perform differential cross section measurements whereby the data are unbinned, possibly in many dimensions. Unbinned measurements can enable, improve, or at least simplify comparisons between experiments and with theoretical predictions.  Furthermore, many-dimensional measurements can be used to define observables after the measurement instead of before.  There is currently no community standard for publishing unbinned data.  While there are also essentially no measurements of this type public, unbinned measurements are expected in the near future given recent methodological advances. The purpose of this paper is to propose a scheme for presenting and using unbinned results, which can hopefully form the basis for a community standard to allow for integration into analysis workflows. This is foreseen to be the start of an evolving community dialogue, in order to accommodate future developments in this field that is rapidly evolving.
}

\maketitle

\clearpage

\section{Introduction}
\label{sec:intro}

One of the main components of high energy particle and nuclear physics research is the prediction and measurement of (differential) cross sections.  Such spectra are sensitive to a wide range of phenomena and can be used to extract fundamental parameters of the Standard Model (SM), explore the quantum properties of the strong and electroweak forces, tune Monte Carlo (MC) event generators, and search for physics beyond the SM.  In order to be comparable to data, predictions must be differential in the measurement observables; in order to be comparable to other experimental data and theoretical predictions, data must be corrected for detector effects.  This is usually done by first discretizing the target phase space into a finite number of bins.  Then, the theoretical and experimental results can be represented as histograms.  These histograms can then be readily visualized and their contents can be encoded in an array for easy storage (e.g.\ in \textsc{HEPData}~\cite{Maguire:2017ypu}) and statistical analysis including MC tuning~\cite{Buckley:2009bj}, Parton Distribution Function (PDF) fits for protons~\cite{Gao:2017yyd} and heavy nuclei~\cite{AbdulKhalek:2020yuc}, Wilson coefficient extractions from higher-dimensional
	operators in the Effective
	Field Theories (EFTs)~\cite{Brivio:2017vri, Biekoetter:2018ypq,daSilvaAlmeida:2018iqo,Brivio:2019ius, Dawson:2020oco, Ellis:2020unq, Ethier:2021ydt, Ethier:2021bye, Brivio:2021alv, Almeida:2021asy}, %
	and global fits of the parameters of UV-complete theories~\cite{GAMBIT:2017snp,Bagnaschi:2015eha} such as supersymmetry.
Correcting for detector effects is known as \textit{unfolding} (see Ref.~\cite{Cowan:2002in,Blobel:2203257,doi:10.1002/9783527653416.ch6,Balasubramanian:2019itp} for recent reviews\footnote{Unfolding is not the only way to publish reusable information about data -- a complementary approach is to publish the full statistical model of the data from parametric fits.  See e.g. Ref.~\cite{2109.04981}.}).

While the current paradigm has resulted in many exciting physics results, there are also significant limitations.   For example, the observables and binning must be chosen ahead of time.  This makes it difficult and in some cases impossible to compare different measurements with each other and with theoretical calculations.  Furthermore, the optimal binning depends on the application.  While there have been some proposals for unbinned measurements in the past~\cite{Lindemann:1995ut,Aslan:2003vu,DEMBINSKI2013410}, until now, they were not used for data analysis.  Recent innovations in machine learning have empowered new algorithms that can process unbinned and high- (or even variable-)dimensional inputs/outputs (e.g. the momenta of all jets in an event)~\cite{Gagunashvili:2010zw,Glazov:2017vni,Datta:2018mwd,Andreassen:2019cjw,Bellagente:2020piv,Bellagente:2019uyp,Andreassen:2021zzk,bunse2018unification,Ruhe2019MiningFS, Howard:2021pos,Vandegar:2020yvw} which are already being used for the analysis of high energy physics data~\cite{2108.12376}.  Representing unbinned spectra is very different than representing binned distributions and therefore it is critically important to discuss common formats and tools to maximize the science potential of the data and the corresponding theoretical predictions.  Our goal is to propose an algorithm-agnostic framework that could become a standard across experiments and theory.  In order to be most useful, widespread consensus is required and so input and feedback is most welcome.

This document is organized as follows. 
First, Sec~\ref{sec:motivation} motivates the use of unbinned measurements.
Then, Sec.~\ref{sec:unbinnedmethods} briefly describes existing proposals for unbinned unfolding methods.  The goal of the proposed format is that it should be method-agnostic and so it needs to be able to accommodate all approaches. 
Section~\ref{sec:statistcs} discusses important statistical properties of unbinned spectra, which will be relevant for ensuring that all necessary information is stored in the proposed format.  The new format itself is described in Sec.~\ref{sec:formatandstorage}.   An example is provided in Sec.~\ref{sec:example}.  The paper ends with conclusions and outlook in Sec.~\ref{sec:conclusion}.  Our proposed format is meant to be the start of an evolving community dialogue, in order to accommodate future developments in this field that is currently rapidly evolving.

\section{Motivation}
\label{sec:motivation}

There are generally three inter-related motivations for unbinned measurements, which are discussed below and combined with motivating physics examples. 

\subsection{Inference-Aware Binning}

 The first and foremost advantage is that measurements which are unbinned can be binned at inference time using the optimal choice of bin boundaries (`inference-aware binning').  Currently, bins are chosen \textit{a priori} without a particular statistical analysis in mind and use heuristic approaches (e.g. bin purity $\geq 50\%$) to achieve a useful binning for a wide range of \textit{post hoc} analyses.  For a given statistical analysis, there will be an optimal choice of binning that minimizes the uncertainty.  For global fits combining multiple analyses, the optimal binning for a particular input measurement could even change with time as other results are updated or added.  Multiple methods have been proposed to determine the preferred binning such as Bayesian reweighting~\cite{Ball:2011gg} and Hessian profiling~\cite{Paukkunen:2014zia} for PDF fits and information geometry techniques for Higgs and EFT physics~\cite{Brehmer:2016nyr}.  One can always attempt a brute-force approach
	whereby a global analysis is repeated
	several times for various choices
	of potential binning, but
	this option is usually disfavoured due to the heavy computational requirements.

The choice of an optimal binning at the time of a particular statistical analysis is facilitated by an unbinned result.  This is particularly true for measurements of multiple moments of distributions (see e.g. Ref.~\cite{CLAS:2009ngd,1509.05190}) and one could even perform an unbinned or neural network-based analysis~\cite{Chen:2020mev,DAgnolo:2018cun,DAgnolo:2019vbw,Brehmer:2018hga,Brehmer:2018kdj,Brehmer:2018eca,Nachman:2019dol} without resorting to bins in the first place.  With the unbinned data, one could study the loss in sensitivity from binning.  In many cases, the information loss will be small (as the binning is adapted to the detector resolution), but the unbinned data can be used to ensure that there is never a significant loss in sensitivity from binning.  This is particularly important for probing the tails of distributions, where any information about the distribution within a bin can improve the sensitivity.  For example, such tails are critical for assessing the interplay between EFT
and PDF fits~\cite{Greljo:2021kvv}.  Ultimately, the detector resolution and its uncertainty set a limit on the amount of information that can be extracted using unbinned spectra.  In some cases, it may not be helpful to consider finer bins, but it could be useful for a variety of reasons to shift bin boundaries. 

The covariance matrices released together with binned unfolded measurements sometimes have very small eigenvalues, translating into instabilities at the level of the $\chi^2$ computations.
The availability of unbinned observables would allow a more detailed investigation and possibly a solution for such problem since it would be possible to rebin without necessarily preserving the previous bin edges (in binned measurements, one can only merge, but not shift bins).

\subsection{Derivative Measurements}

The second advantage of unbinned unfolding is that it enables measurements of derivative observables.  Suppose that one has measured a multi-differential cross section of $x_0, ...,x_n$.  Unbinned methods allow for the \textit{post hoc} measurement of $f(x_0,...,x_n)$ for $f:\mathbb{R}^n\rightarrow\mathbb{R}^m$.  In the binned case, only a crude measurement of $f$ is possible, using the values of $f$ given by the combinations of bin centers over values of the $x_i$.  

A common challenge in global analyses is establishing which observables are maximally sensitive to the parameters one aims to extract from the data. 
	In some cases, one can exploit physical
	arguments in favor of specific options, e.g.\ in EFT fits
	it is often advantageous to bin data in terms
	of some invariant-mass variable to benefit from the energy-growth induced by the EFT operators. 
	Another example is provided by PDF fits, where in general observables that are more closely related to the underlying partonic kinematics (such as lepton rapidity distributions in Drell-Yan production) exhibit a higher sensitivity to the PDFs. 
	However, beyond these rather generic considerations, it is difficult to quantify {\it a priori} which choice of binning or even of differential distribution enjoys the maximal sensitivity.
	Even worse, the answer to this question may vary with time, since as the global fit evolves the relative weight of specific directions in the parameter space vary.

Another important example is the measurement of observables in different reference frames.  For example, in $ep$ scattering, it can be useful to change between the laboratory frame and the Breit frame ($\gamma^*p$ center-of-mass frame).   A seamless change of reference frame would benefit comparisons between different theoretical frameworks, which sometimes are used to study the same final state (e.g. Ref.~\cite{Liu:2018trl} and Ref.~\cite{Gutierrez-Reyes:2018qez}). 

As with binned measurements, it is essential that unbinned measurements include a well-defined phase space (sometimes called the `fiducial volume') for making comparisons to other measurements and to theory predictions.   A benefit of unbinned unfolding is that one can more readily restrict the phase space after the measurement to match another measurement or prediction that has a smaller fiducial volume. 

\subsection{Extension to Higher Dimensions}

For many of the machine learning-based methods, it is straightforward to increase the dimensionality of the unfolding.  While not strictly a benefit of unbinned unfolding directly, there is a close connection between the unbinned nature of machine learning techniques and their ability to accommodate more observables.  This is further enhanced by the considerations in the previous section.  Highly differential binned cross sections are rare since it is cumbersome/impractical to construct and utilize measurements with so many bins. Unbinned measurements provide a practical alternative, but also come with smoothness assumptions related to the interpolations between different phase-space regions, implicitly done in the parameterisation of the detector response used at the unfolding step.

It is possible to take $n$ different one-dimensional measurements and correlate them through bootstrapping for statistical uncertainties\footnote{The nuisance parameters for systematic uncertainties of the same type are assumed fully correlated between measurements.} (see e.g.\ Ref.~\cite{ATL-PHYS-PUB-2021-011} and Ref.~\cite{ATLAS:2017ble} for an example using the inclusive jet and dijet double-differential cross-section measurements in ATLAS), but this requires access to the event numbers in data (and to a lesser extent, Monte Carlo) and the exact analysis selection code.  These are usually only available within a collaboration and if this is not done soon after an analysis is done, it often becomes impossible or at least impractical later.   In contrast, if all relevant dimensions are included in the initial unfolding, then performing a fit to multiple spectra is straightforward.  See e.g.\ the discussion of Refs.~\cite{Czakon:2016olj,Bailey:2019yze,Hou:2019efy} in the case
of PDF analyses using top quark pair production
measurements.

High-dimensional measurements are particularly important for probing hadronic final states, where it is important to simultaneously probe multiple objects.  For example, studies of azimuthal modulations in semi-inclusive deep inelastic scattering (SIDIS) and related processes such as hadron-in-jet measurements need to include both electron and hadron variables~\cite{Arratia:2020nxw}.  Such measurements often require unfolding $\mathcal{O}(10)$ dimensions at once, which is challenging with binned approaches.  It is known that limited geometrical acceptance of detectors introduce coupling between different harmonics, and those effects are compounded when performing binned measurements, especially when integrating over kinematic variables, see e.g.\ Ref.~\cite{Hayward:2021psm}.   Furthermore, the detector resolution can depend on many quantities, so simultaneously including those observables will be necessary to achieve the ultimate precision.  


All in all, there are strong reasons
why unbinned measurements would represent
a breakthrough in the way particle and nuclear physics measurements at the LHC and elsewhere
are both presented and used in theoretical
interpretations, which would be specially timely
with the upcoming era of the high-statistics
High Luminosity LHC~\cite{Azzi:2019yne,Cepeda:2019klc} and the electron ion collider (EIC)~\cite{AbdulKhalek:2021gbh}.


\section{Overview of Unbinned Methods}
\label{sec:unbinnedmethods}

As the goal of the proposed format is to be unfolding-procedure agnostic, it is important to briefly introduce current proposals so that the resulting format can be universally accommodating.  This section is not meant to be comprehensive -- please see the original papers for further details.   Additional examples of generative models or classification models in high energy physics can be found in Ref.~\cite{2102.02770}.

\subsection{Experimental Measurements}

For the remainder of this section, each data event will be represented by $x\in\mathbb{R}^n$, where $n$ could be one for a single target observable or $n>1$ for a multidimensional unfolding.  The dimensionality of the detector-level and particle-level observables need not be the same and unless specified, $x$ will refer to the particle-level features.

\subsubsection{Density-Based Models}
\label{sec:density}


Approaches of this type learn an implicit or explicit function $\hat{p}(x):\mathbb{R}^n\rightarrow\mathbb{R}$ that represents the probability density of the unfolded result that is designed to estimate the true density $p(x)$.  Low-dimensional approaches without neural networks have been proposed based on non-parametric density estimators~\cite{Dembinski:2013hdz}. 
 Deep generative models like Generative Adversarial Networks (GAN)~\cite{Goodfellow:2014:GAN:2969033.2969125,Creswell2018} and Variational Autoencoders (VAE)~\cite{kingma2014autoencoding,Kingma2019} produce implicit models for $\hat{p}$, i.e. they can sample from $\hat{p}$ but do not provide a method for evaluating it explicitly.  In contrast, deep models based on Normalizing Flows (NF)~\cite{10.5555/3045118.3045281,Kobyzev2020} allow for both sampling and density estimation.  GAN-based proposals include Ref.~\cite{Datta:2018mwd,Bellagente:2019uyp}, Ref.~\cite{Howard:2021pos} relies on a VAE, and Ref.~\cite{Bellagente:2020piv,Vandegar:2020yvw} proposed using NFs.

\subsubsection{Classifier-Based Models}
\label{sec:classifier}

Approaches of this type learn a function $\hat{w}(x)$ that approximates the likelihood ratio $w(x)= p(x)/p_\text{MC}(x)$, where $p(x)$ is the true probability density of the unfolded data as discussed in the previous subsection, and $p_\text{MC}$ is the equivalent quantity predicted by a Monte Carlo algorithm. With this function one can sample events from the nominal MC and weight each event by $\hat{w}(x_i)$, where $x_i$ are the features of the event.  The produced sample has now been reweighted to match the unfolded data and constitutes the (central value) of the unbinned measurement.
Any expectation values computed using this sample can be used as the central value of a measurement of this quantity, and if we include a normalization factor as part of $\hat{w}(x)$ that account for the integrated luminosity of the dataset, the sum of weights of any kinematic region defined from $x$ will correspond to the associated fiducial cross section. Learning $\hat{w}$ is achieved by making use of a well-known~\cite{hastie01statisticallearning,sugiyama_suzuki_kanamori_2012} fact that one can directly learn likelihood ratios using classifiers, e.g. $\hat{w}(x)=f(x)/(1-f(x))$ for classifier $f(x)$ trained with the cross-entropy loss function.  This turns the problem from density estimation (estimating $p(x)$ and $p_\text{MC}(x)$ separately) into classification.   Classifier-based proposals include Ref.~\cite{Andreassen:2019cjw,Andreassen:2021zzk,bunse2018unification,Ruhe2019MiningFS}.

\subsection{Theoretical Calculations}

While the main motivation for the format proposed in this report is experimental measurements, it can also be used for theory predictions.  In fact, we strongly encourage the publication of theoretical predictions alongside experimental measurements.
If the latter are presented unbinned, then the predictions should also be presented unbinned\footnote{Sometimes, predictions are complemented by weights that represent corrections for higher-orders (`$k$-factors') or for non-perturbative effects.  These could also be unbinned (and maybe could use machine learning for parameterization in a given kinematic feature like jet $p_T$).}.  Unbinned predictions could also be provided before a measurement is done, without the need to iterate between theorists and experimentalists on the exact analysis binning.  Related ideas for unbinned theory predictions have been discussed in the context of (next-to)next-to leading order $n$-tuples (see Ref.~\cite{Heinrich:2016jad} and references therein) whereby all or part of a calculation can be stored for later reuse.  Just like for experimental results, theoretical predictions come with uncertainties.  Some of these uncertainties can be computed `on the fly' (e.g. in parton showers~\cite{Mrenna:2016sih,Bellm:2016voq,Bothmann:2016nao} and parton densities~\cite{Buckley:2014ana}) if certain information like the colliding particle types and energies are stored.  This information is not observable (in machine learning, this is called `latent') and so would not have a corresponding entry in an unbinned measurement.  While it could be interesting to store such information for on-the-fly calculation of uncertainties, we propose to focus on fully prepared results that do not need any additional machinery for interpretation.  This could be revisited in the future.  However, it is necessary to store the output of the weight variations so that uncertainties can be computed, even if the weights cannot be rederived from the available (observable) information.

\section{Statistics of Unbinned Spectra}
\label{sec:statistcs}

\subsection{Acceptance Effects}

An essential aspect of unfolded spectra is that they should not depend on detector-level quantities.  This means that in addition to correcting for resolution effects, unfolding algorithms must also account for finite acceptance effects.  One way to do this unbinned was proposed in Ref.~\cite{Andreassen:2021zzk}: simply assign an `empty' event symbol to events that pass the detector-level or particle-level event selection (but not both).  Particle-level empty events are discarded from the final result.  One should also be cautious of matching objects between detector-level and particle-level because this can introduce a reliance on the detector-level objects.  For example, if one measures properties about the leading jet in an event, it is important to not use a $p_T$-matching to pick the particle-level jet -- often is is better\footnote{It may instead be better to use a geometric matching.  That renders the measurement dependent on the detector-level observables, but given the superb angular resolution of modern detectors, this may be a negligible effect.  Another approach is to use a different scheme altogether (e.g. rapidity ordering~\cite{1509.05190}).} to use the leading particle-level jet independent of how close it might be to the leading detector-level jet.

\subsection{Background Subtraction}

Many measurements have contamination from background processes that must be statistically subtracted to produce a measurement in the desired region of particle-level phase space.  For example, a measurement involving charged leptons may need to estimate and subtract the contribution from events with \textit{fake} charged leptons.  When measuring binned spectra, this can be done on a bin-by-bin basis.  There are multiple unbinned generalizations.  For density-based methods (Sec.~\ref{sec:density}), one can learn a subtracted density~\cite{Butter:2019eyo}.  For classifier-based approaches (Sec.~\ref{sec:classifier}), one can assign a negative weight to events that are to be statistically subtracted.  If one wants to avoid negative weights, unbinned positive reweighting is possible~\cite{Nachman:2020fff}.  The resulting weighted events could also be turned into unweighted events, possibly based also on neural networks~\cite{Backes:2020vka, Stienen:2020gns}.

Many backgrounds are estimated using data-driven methods like the fake factor and ABCD methods.  These approaches are usually constructed with bins, but unbinned generalizations are possible.

\subsection{Local Statistical Uncertainty}

To compute the expectation value of an observable from $N$ weighted events, one needs to average that observable over the event weights:
\begin{equation}
    \label{eq:weight_sum}
    \hat{\mu}_\mathcal{O} = \sum_{i=1}^N w_i \, \mathcal{O}(x_i),
\end{equation}
where $w_i$ are the individual event weights
and $\mathcal{O}$ is the observable on interest that depends on the event features $x_i$.
For example, in the case of a binned differential cross section, the observable would be the indicator function for a histogram bin (0 if the event lands outside of the bin of interest, 1 if it lands inside).  The functions $\mathcal{O}$ and/or the weights $w$ may be dimensionful depending on the application.  For example, in the case of a binned differential cross section, $\hat{\mu}$ will have units of cross section divided by the units of the differential quantity.  When measuring a summary statistic like the average of a quantity, the weights $w_i$ are normalized to unity.

In the large $N$ limit, the standard way to compute statistical uncertainties on $\hat{\mu}_\mathcal{O}$ is to sum over the squared weights:
\begin{equation}
    \label{eq:weight_squared_sum}
    \delta\hat{\mu}_\mathcal{O} = \sqrt{\sum_{i=1}^N w_i^2 \, \mathcal{O}(x_i)^2}.
\end{equation}
This corresponds to estimating the standard error of the mean and dropping terms that are higher-order in $N$.
For equal weight events and a histogram-based observable, this yields the standard result that $\delta\hat{\mu}_\mathcal{O} / \hat{\mu}_\mathcal{O} = 1/\sqrt{N_\mathcal{O}}$, where $N_\mathcal{O}$ is the number of events in the histogram bin.

Because \Eqs{weight_sum}{weight_squared_sum} depend on the same $w_i$, one can derive the standard statistical uncertainty by making sure each event is stored with its own weight.
There are contexts, however, where the standard formula does not apply.
For example, if there is a large variability in the event weights or if there are negative weight events, it may be desirable to reweight the events in a local phase space patch while preserving the cross section of that patch~\cite{Andersen:2020sjs}:
\begin{equation}
    \label{eq:weight_patch}
    w_{\rm patch} = \frac{1}{N_{\rm patch}} \sum_{\text{events $i$ in patch}}^{} w_i.
\end{equation}
For small enough patches, this replacement preserves the expectation value in \Eq{weight_sum}.
As emphasized in \Ref{Nachman:2020fff}, however, it does not preserve the uncertainties:
\begin{equation}
    \label{eq:weight_squared_patch}
    w_{\rm patch}^2 \not= \frac{1}{N_{\rm patch}} \sum_{\text{events $i$ in patch}}^{} w_i^2.
\end{equation}

The solution advocated in \Ref{Nachman:2020fff} was to resample events, such that a fraction of events are removed and the remaining events are upweighted such that both \Eqs{weight_sum}{weight_squared_sum} are preserved.
This solution might not be desirable in all circumstances, though, so we advocate for separately storing
\begin{equation}
w_i, \qquad \tilde{w}_i^2,
\end{equation}
where $\tilde{w}_i^2$ has the interpretation as the local average squared weight.
In the reweighting example above, one would simply store the right hand sides of \Eqs{weight_patch}{weight_squared_patch}.
In situations where this extra information is not needed (i.e.\ there is no resampling), a value of $\tilde{w}_i^2 = w_i^2$ should be used. %

\section{Systematic Uncertainties}
\label{Sec:systematics}

Typically, systematic uncertainties are estimated by varying some aspect of the simulation and repeating the unfolding procedure.  Such uncertainties can be classified into two-point and replica-type uncertainties (discussed more in Sec.~\ref{sec:uncerts}). This section briefly discusses uncertainties related to the unfolding procedure itself, which may be more involved when using machine learning.

Potential biases from the unfolding procedure are typically evaluated through a set of non-closure tests.  In particular, the unfolding procedure is repeated for a case where the answer is known and the difference between the unfolded result and the known answer can be used to set a systematic uncertainty on the method bias.  Examples of this type include unfolding one generator (labeled `data') with another generator (labeled `simulation') as well as stress-tests whereby one generator (labeled `data') is deformed in some way and it is then unfolded with the same, non-deformed generator (labeled `simulation').  
For example, a stress test may be created by weighting each event so that the ratio between the weighted and unweighted distribution is linear with a particular slope.  
As with all systematic uncertainties, there is no unique way to determine the non-closure uncertainty and ideally the data should be used in some way to bound the range of reasonable variations studied.
Indeed, the method proposed in Ref.~\cite{Malaescu:2009dm} allows to perform a data-driven closure test, introducing weights at particle-level so that the corresponding detector-level distribution is a better match to the data. This approach has the advantage of emulating the conditions of the actual data unfolding, hence its reliability.
In general, several non-closure tests have to be performed and their output synthesized; and one has to make sure that the tests probe complementary effects, in order to avoid a possible double counting of some effects between such tests.

The same non-closure uncertainties discussed above can be used in the unbinned case.  No additional uncertainties are required simply because neural networks are used.  However, the architecture and training of neural networks may lead to sub-optimal performance and it may be useful to estimate the potential deviation from optimal performance (referred to as `optimally uncertainties' in Ref.~\cite{Nachman:2019dol}) to ensure that they are small.  In particular, density-based methods (Sec.~\ref{sec:density}) assume that the learned density is the true density and classifier-methods (Sec.~\ref{sec:classifier}) assume that the learned function is a monotonic transformation of the true likelihood-ratio.  If `optimally uncertainties' are not small, then the architecture/learning could be improved or additional steps could be added to enhance accuracy (such as calibration~\cite{Cranmer:2015bka}).  There are multiple ways to probe these uncertainties, including reinitalizing with different starting parameters, bootstrapping, etc.  Bayesian neural networks may also be able to help quantify some of these `uncertainties'~\cite{Bellagente:2021yyh}.

\section{Proposed Format and Storage}
\label{sec:formatandstorage}

\subsection{Structure}

There are generically two ways to represent unbinned data.  If the results can be encoded by a parameterized probability density, then one could simply publish the functional form and fitted parameters.  A second approach would be to publish data sampled from the unbinned result.  This second approach is more general as not all methods produce a fitted function (see Sec.~\ref{sec:unbinnedmethods}) and therefore it is the baseline for our proposal\footnote{In the future, there may also be interest in unfolding \textit{parameterized observables} such that the observables themselves depend on an input $\mu$.  This could be done by providing values sampled in $\mu$, but this could result in some loss of information depending on how the unfolding is performed.  Exploring this possibility is left for future work.}.  However, we strongly encourage publishing any other representation as well, including neural network architectures (with fitted weights and biases) which can be used to either generate new samples (Sec.~\ref{sec:density}) or reweight new samples (Sec.~\ref{sec:classifier}).  Neural networks would ideally be converted to a standard format like the Open Neural Network Exchange (ONNX)~\cite{bai2019} or the native format if a given architecture is not supported in ONNX or precision is lost.  If converting to a non-native format, it is important to verify that there is no loss in precision, which may be especially important for likelihood ratio approximations.

The proposal is to have two files: (1) a \texttt{submission} file, which has the same format as the \texttt{submission} YAML file used for \textsc{HEPData} and (2) a data file for each measurement.  We encourage analyzers to include a default binning with the corresponding histogram contents for easy comparison and validation.  The location of the data files are in the \texttt{submission} file.  Note that a given analysis may include multiple measurements; also a given measurement could have multiple dimensions and from it, multiple observables could be constructed.  Suppose that a given measurement has $k$-dimensions per event.  Further suppose that there are $N$ events sampled from the unfolded result and there are $M$ systematic uncertainties.  Then, the data files for the measurement will be a matrix of size $(M+1)\times (k+2)\times N$.  A common block is repeated for the nominal and all uncertainties.  This block has $N$ columns and $k+2$ rows.  The first $k$ rows are simply the values of the observable $\vec{x}_i\in\mathbb{R}^k$.  The next row encodes per-event weights and the last row encodes the per-event statistical uncertainty.  In most cases, the per-event statistical uncertainty will simply be the per-event weight squared, but it need not be the case (see Sec.~\ref{sec:statistcs}).  The structure of this block is illustrated below, for an algorithm that does not use event weights (top) and one that does (bottom):\\

 \begin{tcolorbox}
\textbf{Example: density-based approach (no event-weights)}\\
$\vec{x}_1$, $\vec{x}_2$, ..., $\vec{x}_N$\\
1, 1, ..., 1\\
1, 1, ..., 1
 \end{tcolorbox}
 
  \begin{tcolorbox}
\textbf{Example: classifier-based approach}\\
$\vec{x}_1$, $\vec{x}_2$, ..., $\vec{x}_N$\\
$w_1$, $w_2$, ... , $w_N$\\
$w_1^2$, $w_2^2$, ... , $w_N^2$
 \end{tcolorbox}
 
 Note that the above tables are represented as one event per column to fit on the page, but we envision that actually each event will be represented as a row (the transpose of the above tables).  Furthermore, for classifier-based methods, for systematic variations that use the same particle-level events, but with different detector-level events or different weights, it could be possible to use the same feature list with a different set of unfolded weights for each uncertainty.  In this case, only $(k+2+2M)\times N$ numbers are required (or even fewer if the local uncertainty is $w^2$).
 
 \subsection{Uncertainties}
 \label{sec:uncerts}
 
 There are two types of uncertainties: two-point and replica.  Two-point uncertainties result from modifying a discrete parameter in the simulation and then rerunning the unfolding.  In this case, there will be an entry in the output that has the same structure as the nominal result, but with different values (either different $\vec{x}_i$, different weights, or both).  When the unfolded data are assembled into a bin, the uncertainty on the bin content is computed by taking the difference in the bin content when using the nominal entry and the varied entry.  
 
 Replica uncertainties are constructed by sampling many unfolded results from a distribution.  They are described by not one entry, but many entries and the uncertainty when binned is computed by taking the standard deviation across values.  The most common replica uncertainties are due to statistical uncertainties where the data or simulation are bootstrapped and the unfolding is repeated a number of times.  Replica uncertainties are also encountered for resolution uncertainties where additional smearing may be added to a reconstructed object many times to determine the impact of an increased resolution.
 
\subsection{Storage}
 
Unlike recording histograms, storing unbinned data in the proposed format may require significant storage space ($\mathcal{O}$(GB)).  We therefore propose to host the results on \textsc{Zenodo} (\url{https://zenodo.org}), but linked to \textsc{HEPData} to make it easier to search and integrate with Rivet routines~\cite{Buckley:2010ar}.

The access to the content of these large files can be streamlined to allow exploration of the data on the server while downloading only a relevant sub-region of the data. This would allow a user to download only the relevant observables from multiple files in an automated fashion and perform a combined analysis. The HDF5 format~\cite{hdf5} allows for querying, and the HDF Server (h5serv)~\cite{h5serv} defines universal unique identifiers (UUID) for each group and dataset object, which means that the user does not require prior knowledge of the HDF5 path in order to access a group in the file. Alternate implementations such as with XRootD~\cite{xrootd} if using root format files or with Paquet~\cite{parquet} and Arrow~\cite{arrow} are also possible. The additional effort to add server side support required for these implementations is relatively small and will become worthwhile as presenting unbinned data becomes widespread.

\subsection{Common Tools}

The integration with standard analysis statistical (e.g.\ \textsc{RooUnfold}~\cite{adye2011unfolding}) and preservation tools (e.g.\ \textsc{Rivet}~\cite{Buckley:2010ar}) will depend on and benefit from the solution chosen for storage and presentation of the unfolded result.  This would require extensions of common tools, such as a machine learning interface to \textsc{RooUnfold} and a utility in \textsc{Rivet} to rebin data on the fly.  However, given the widespread use of these tools for comparing with predictions, it will be important to add to these packages as unbinned measurements become popular.  Additional harmonization would also be useful, such as the standardization of names for various nuisance parameters.  While this is not strictly a requirement of unbinned unfolding, it would facilitate easier comparisons and combinations with existing and future measurements.

\section{Example}
\label{sec:example}

As an example, we consider the $Z+\mathrm{jets}$ production process. The main purpose of this section, is to show in practice the differences of density-based and classifier-based models. As a classifier-based model we utilize the \textsc{OmniFold}~\cite{Andreassen:2019cjw} approach which is an unbinned and multi-dimensional version of the well-established IBU~\cite{DAGOSTINI1995487, 1974AJ79745L,Richardson:72} method. As a density-based model we employ a conditional invertible neural network (cINN) as being used in Ref.~\cite{Bellagente:2020piv} which directly learns a stochastic mapping from detector-level to particle-level.  The code to reproduce both methods can be found at \href{https://github.com/ramonpeter/UnbinnedMeasurements}{this Github repository}.

We use the same simulations as in Ref.~\cite{Andreassen:2019cjw}, which are briefly summarized in the following.   Proton-proton collisions are simulated at $\sqrt{s}=14$ TeV using \textsc{Pythia}~8.243~\cite{Sjostrand:2007gs,Sjostrand:2006za,Sjostrand:2014zea} with Tune 26~\cite{ATL-PHYS-PUB-2014-021}.  Detector effects are emulated using the \textsc{Delphes}~3.4.2~\cite{deFavereau:2013fsa} fast simulation of the CMS detector, which uses particle flow reconstruction.  The \textsc{Pythia} simulation will be called `data' and is also the model for the nominal `simulation'.  An alternative simulation, to illustrate a systematic uncertainty, is provided by \textsc{Herwig}~7.1.5~\cite{Bahr:2008pv,Bellm:2015jjp,Bellm:2017bvx} with its default tune.  Jets are clustered using the anti-$k_t$ algorithm~\cite{Cacciari:2008gp} with radius parameter $R=0.4$ as implemented in \textsc{FastJet}~3.3.2~\cite{Cacciari:2011ma,Cacciari:2005hq}, using either  all particle flow objects (detector-level) or stable non-neutrino truth particles (particle-level).  To reduce acceptance effects, the leading jets are studied in events with a $Z$ boson with transverse momentum $p_T^Z>200$~GeV.  A set of $500,000$ events are set aside as `data' and $500,000$ events from both \textsc{Pythia} and \textsc{Herwig} are then used for the simulation.  To keep this example focused, we consider a two-dimensional unfolding of the $N$-subjettiness variables $\tau_1$ and $\tau_2$~\cite{Thaler:2010tr,Thaler:2011gf}.

\begin{figure}[h!]
    \centering
    \includegraphics[width=0.7\textwidth]{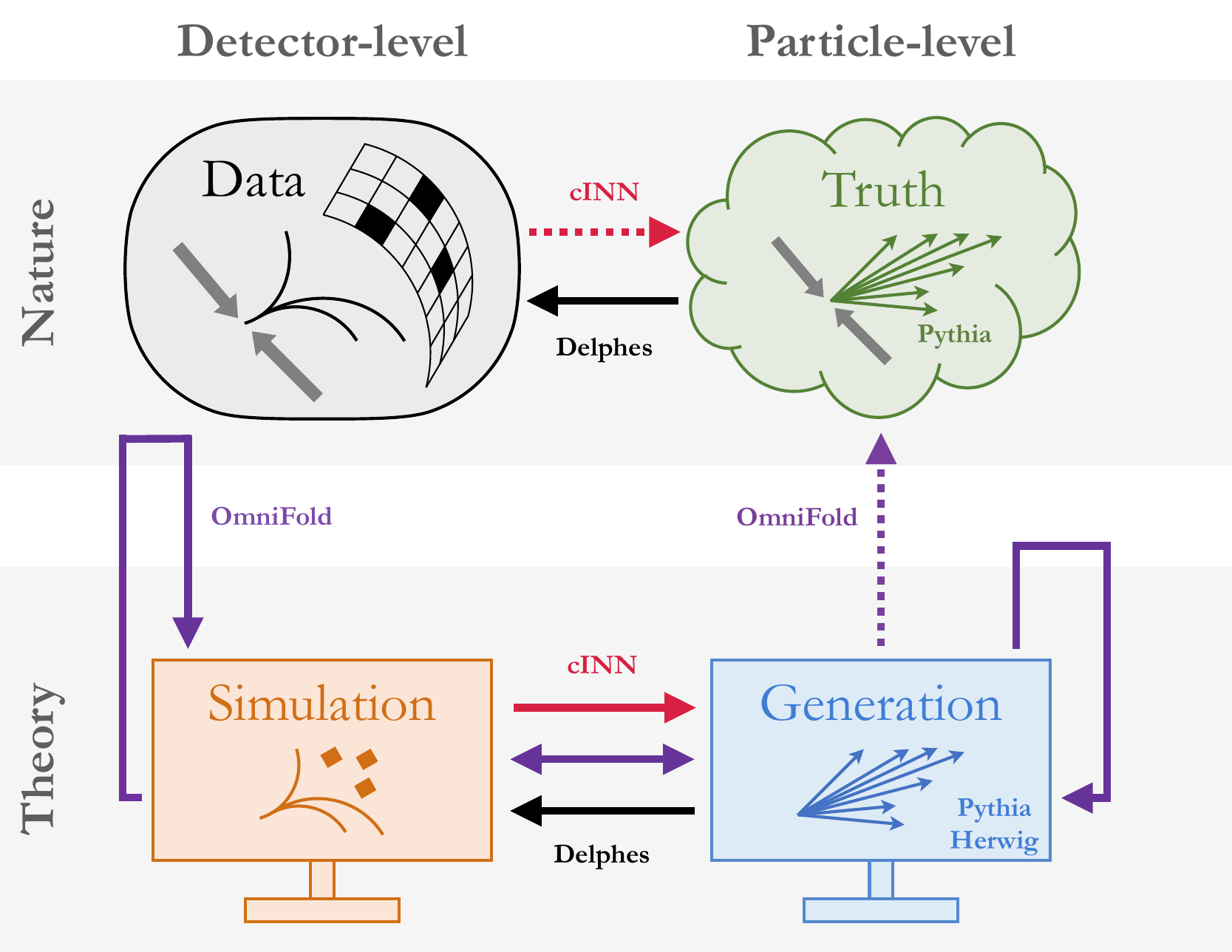}
    \caption{An schematic overview illustrating which parts are connected via several different models and tools.
    The black arrows indicate the usage of \textsc{Delphes}. The red arrows show which parts are connected by the cINN method. In contrast, the iterative procedure of \textsc{OmniFold} is highlighted by the purple arrows.  Adapted from Ref.~\cite{Andreassen:2019cjw}.}
    \label{fig:overview}
\end{figure}

Both neural network models are implemented in \textsc{TensorFlow}~\cite{tensorflow} and optimized using \textsc{Adam}~\cite{adam}.  The \textsc{OmniFold} network consists of three hidden layers with 50 nodes per layer using the rectified linear activation function (ReLU) for intermediate layers and a sigmoid for the final layer.   Training proceeds for 20 epochs with a batch size of 10,000 and the algorithm is repeated for five iterations. The cINN stacks 12 affine coupling blocks, where each block consists of a fully connected multi-layer perceptron with 4 hidden layers, 16 nodes per layer and leaky ReLU as activation function. It is trained for 50 epochs and a batch size of 1000.  None of these parameters were optimized. An schematic overview comparing both employed models is illustrated in Fig.~\ref{fig:overview}.

In the case of the cINN, the nominal result is built by training on the particle-level and detector-level `simulation' and then applied to the detector-level `data'.  Ten particle-level events are sampled from the density for each detector-level `data' event.  In order to preserve the total cross section, each event is then given a weight of $w_i=0.1$.  The first five sampled events in the proposed format are shown below ($\vec{x}_i=(\tau_{1,i},\tau_{2,i})$):

\vspace{5mm}

 \begin{tcolorbox}
\textbf{cINN: Nominal}\\
(0.093,0.067), (0.138,0.079), (0.095,0.066), (0.056,0.054), (0.362,0.261) , ... \\
0.1, 0.1, 0.1, 0.1, 0.1 , ... \\
0.01, 0.01, 0.01, 0.01, 0.01 , ...
 \end{tcolorbox}
 
\noindent The systematic uncertainty due to the MC model is estimated similarly.  In particular, the cINN is trained using the particle-level and detector-level samples from the alternative simulation.  This model is then applied to the `data', where ten events are sampled from each `data' event.  The first five samples are shown here:
 
 \vspace{5mm}
 
  \begin{tcolorbox}
 \textbf{cINN: Systematic Uncertainty (Alternative MC)}\\
(0.059,0.042), (0.150,0.102), (0.117,0.086), (0.090,0.065), (0.091,0.050) , ... \\
0.1, 0.1, 0.1, 0.1, 0.1 , ... \\
0.01, 0.01, 0.01, 0.01, 0.01 , ...
 \end{tcolorbox}
 
  \vspace{2mm}
 
\noindent For the nominal OmniFold result, both the `data' and `simulation' are used to derive weights.  Since the `data' and `simulation' are statistically identical, the weights are nearly constant and unity, as shown below:
 
 \vspace{5mm}
 
 \begin{tcolorbox}
\textbf{OmniFold: Nominal}\\
(0.070,0.071), (0.206,0.181), (0.117,0.053), (0.292,0.122), (0.029,0.027) , ... \\
1.003, 1.003, 1.003, 1.003, 1.003 , ... \\
1.007, 1.007, 1.007, 1.007, 1.007 , ...
 \end{tcolorbox}
 
  \vspace{2mm}
 
\noindent Finally, the alternative MC OmniFold result is shown below.  Now that the data and corresponding simulation are not statistically identical, the weights follow a non-trivial distribution:
 
 \vspace{5mm}
 
  \begin{tcolorbox}
 \textbf{OmniFold: Systematic Uncertainty (Alternative MC)}\\
(0.185,0.086), (0.063,0.065), (0.400,0.156), (0.179,0.125), (0.050,0.047) , ... \\
0.696, 1.626, 0.658, 0.777, 1.713 , ... \\
0.485, 2.645, 0.433, 0.604, 2.934 , ...
 \end{tcolorbox}

  \vspace{2mm}
  
\begin{figure}[h!]
\centering
\includegraphics[width=0.6\textwidth]{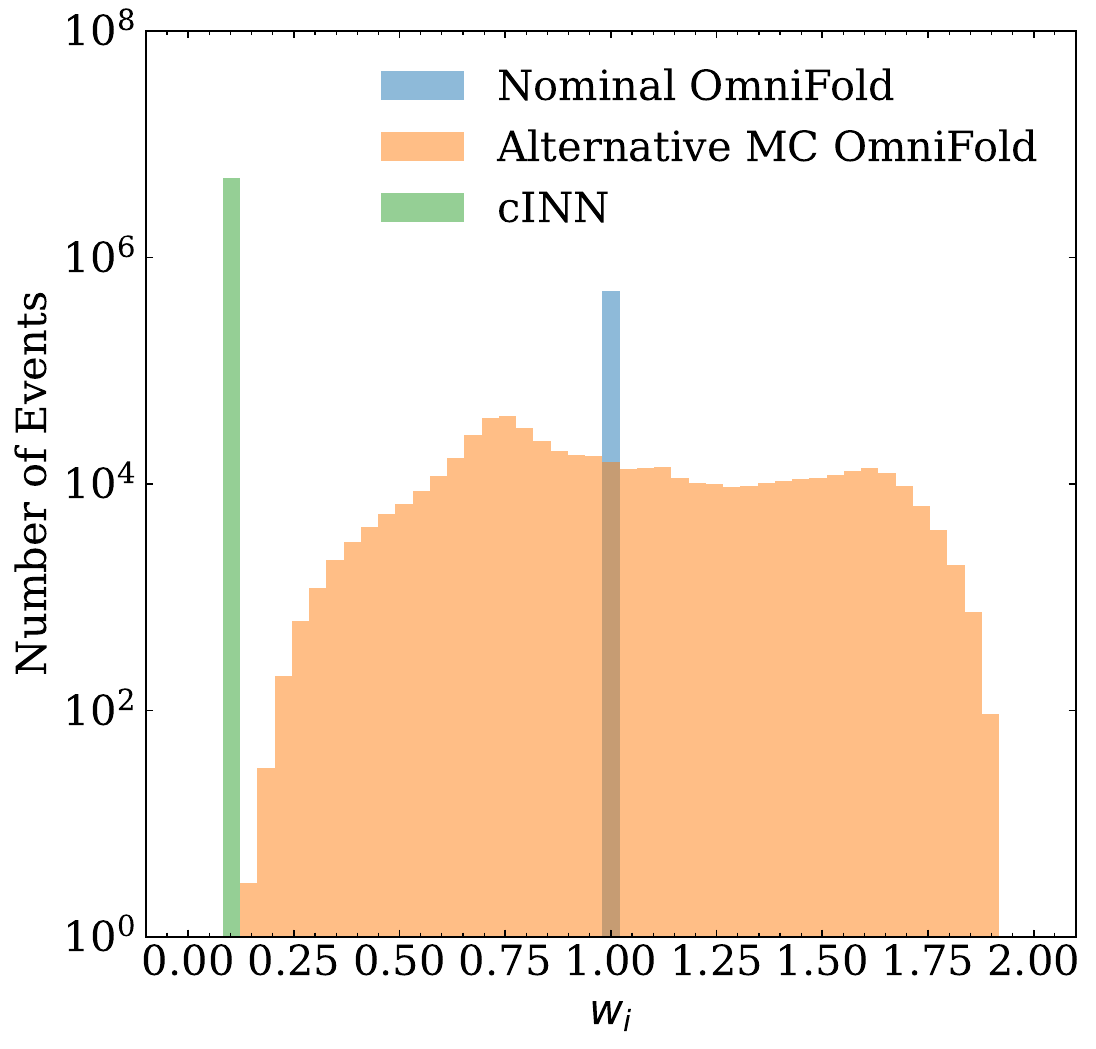}
  \caption{A histogram of the weights for each method.  By definition, the weights for the cINN are 0.1, since 10 events are sampled from the density for each event in `data'.  There are correspondingly ten times more events for the cINN than for OmniFold.  For OmniFold, the nominal weights are nearly all one since the data and simulation are statistically identical.}
    \label{fig:weights}
\end{figure}

\begin{figure}[h!]
\centering
\includegraphics[width=0.49\textwidth]{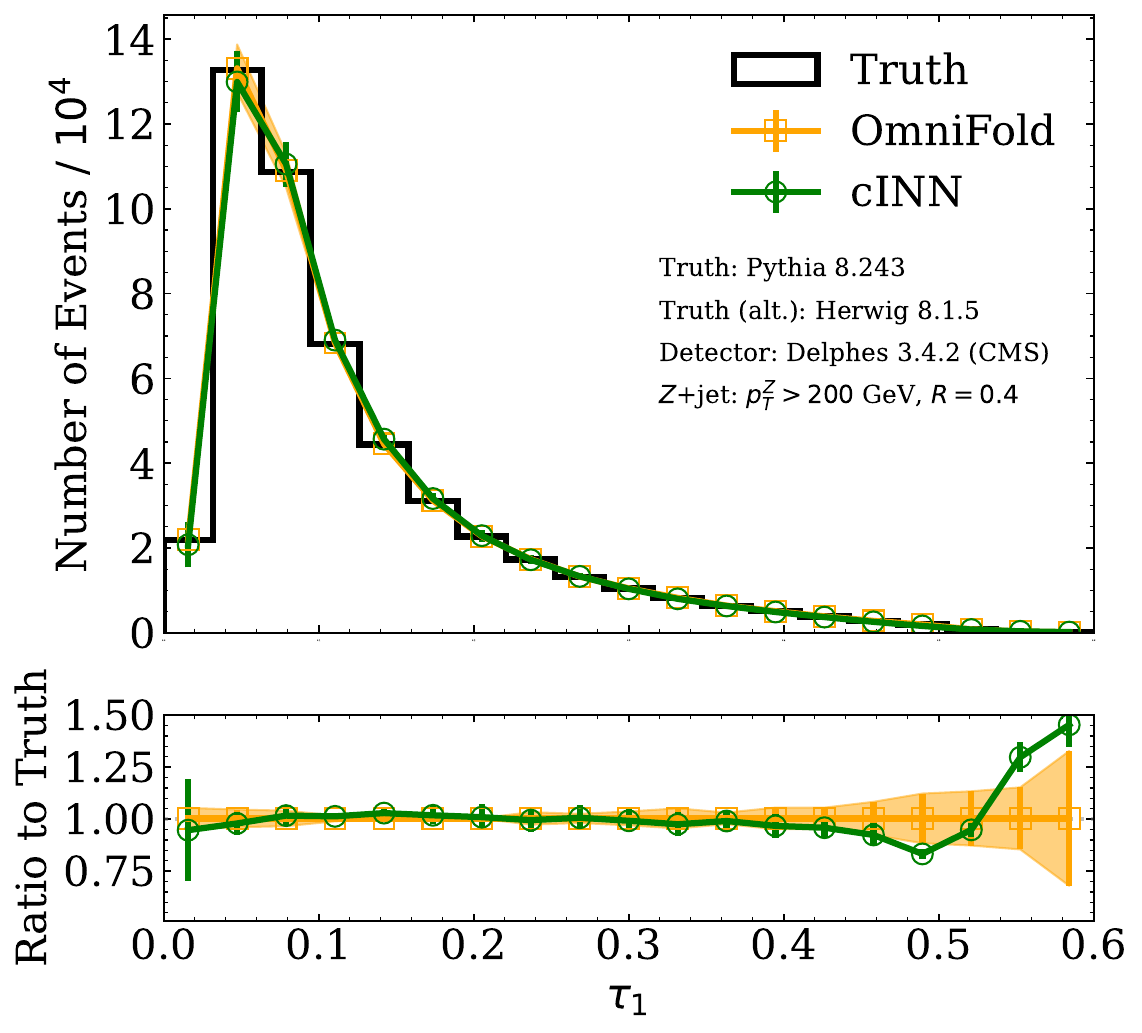}\includegraphics[width=0.49\textwidth]{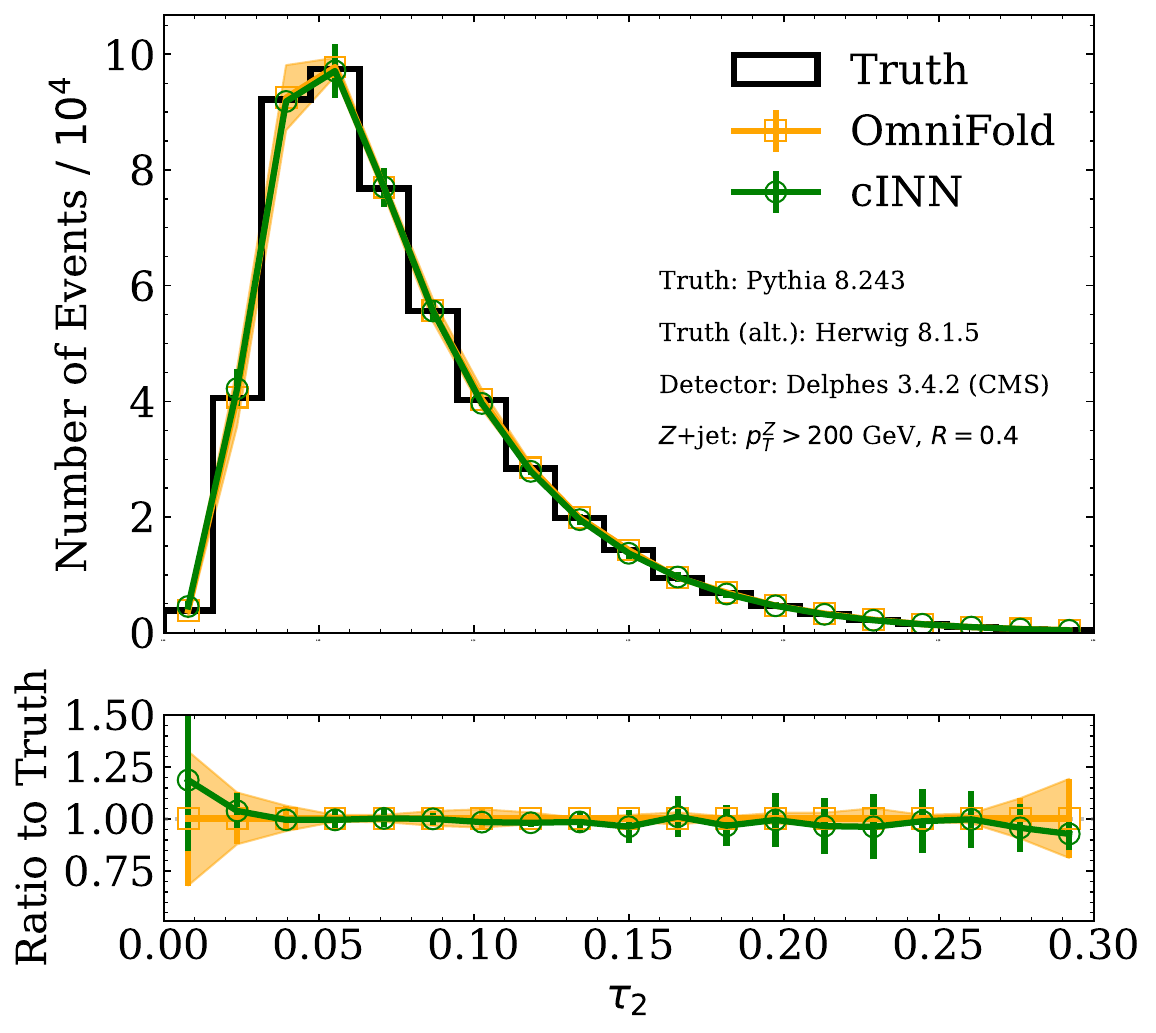}
  \caption{The measured histograms of $\tau_1$ (left) and $\tau_2$ (right) using \textsc{OmniFold} and the cINN methods.  The error band represents the modeling uncertainty from comparing \textsc{Pythia} and \textsc{Herwig}.}
    \label{fig:dists}
\end{figure}

\begin{figure}[h!]
\centering
\includegraphics[width=0.8\textwidth]{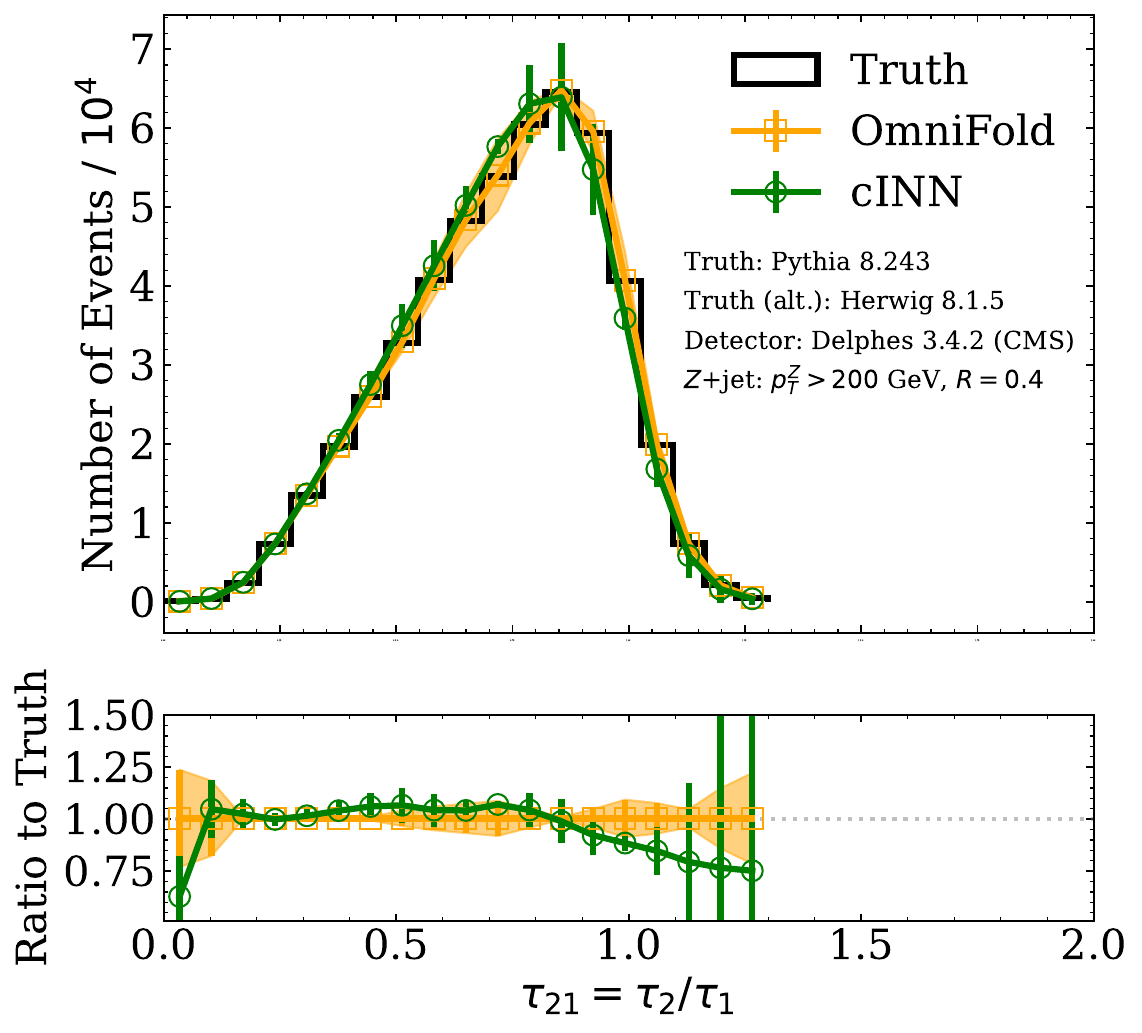}
  \caption{The measured histogram of $\tau_{21}=\tau_2/\tau_1$ using \textsc{OmniFold} and the cINN methods.  The error band represents the modeling uncertainty from comparing \textsc{Pythia} and \textsc{Herwig}.}
    \label{fig:ratio}
\end{figure}

  A histogram of the weights for all methods is presented in Fig.~\ref{fig:weights} and the corresponding unfolded distributions, with a systematic uncertainty band is shown in Fig.~\ref{fig:dists}.  As advertised in Sec.~\ref{sec:motivation}, a benefit of unbinned unfolding is that it is easy to define new observables.  This is demonstrated in Fig.~\ref{fig:ratio}, which shows unfolded results for $\tau_{21}=\tau_2/\tau_1$, which can be derived from the two-dimensional unfolded phase space.  The OmniFold and cINN results are comparable; as stated earlier, neither method was optimized and it would be exciting to quantitatively compare the two approaches in the future.

\clearpage

\section{Conclusions and Outlook}
\label{sec:conclusion}

This report has presented a proposal for publishing unbinned differential cross section measurements and predictions.  The methodology for unbinned measurements has been enhanced in part because of recent advances in machine learning.  Unbinned measurements can always be rebinned and analyzed with traditional methods.  In order to make best use of future measurements that use these tools, it is important to have some community standards and guidelines.  We hope that this paper will serve the purpose of having a concrete proposal to discuss with all of the relevant experimental and theoretical communities.  This is foreseen to be the start of an evolving community dialogue, in order to accommodate future developments in this field that is currently rapidly evolving.

\section*{\label{sec::acknowledgments}Acknowledgments}

AB is supported by the DFG (German Research Foundation) under grant 396021762 – TRR 257.
BM gratefully acknowledges the continuous support from LPNHE, CNRS/IN2P3, Sorbonne Universit\'e and Universit\'e de Paris.
BN was supported by the Department of Energy, Office of Science under contract number DE-AC02-05CH11231.
JR is partially supported by NWO, the 
Dutch Science Council.
JT is supported by the National Science Foundation under Cooperative Agreement PHY-2019786 (The NSF AI Institute for Artificial Intelligence and Fundamental Interactions, \url{http://iaifi.org/}), and by the U.S. DOE Office of High Energy Physics under grant number DE-SC0012567.
RW is supported by FRS-FNRS (Belgian National Scientific Research Fund) IISN
projects 4.4503.16.
We would like to thank the organizers of the Les Houches PhysTeV workshop series (especially Emanuele Re), which hosted the online forum that began this document in June of 2021.

\bibliographystyle{JHEP}
\bibliography{main,HEPML}

\end{document}